\documentstyle[prd,preprint,aps]{revtex}
\newcommand{\be}{\begin{equation}}
\newcommand{\ee}{\end{equation}}
\newcommand{\bea}{\begin{eqnarray}}
\newcommand{\eea}{\end{eqnarray}}

\newcommand{\JC} {J_{cl}}
\newcommand{\JA} {J_{\alpha}}
\newcommand{\JG} {J_{\gamma}}
\newcommand{\JB} {J_{\beta}}
\newcommand{\PAA} {P_{A}}
\newcommand{\PC} {P_{cl}}
\newcommand{\PQ} {P_{Q}}
\newcommand{\PG} {P_{\gamma}}
\newcommand{\PA} {P_{\alpha}}
\newcommand{\JX} {J_{x}}
\newcommand{\JY} {J_{y}}
\newcommand{\JZ} {J_{z}}

\newcommand{\IM} {\Pi_{M}}
\newcommand{\IC} {\Pi_{cl}}
\newcommand{\IQ} {\Pi_{Q}}
\newcommand{\IA} {\Pi_{\alpha}}
\newcommand{\IG} {\Pi_{\gamma}}
\newcommand{\dd} {\partial}
\newcommand{\CX} {{\cal P}_{X}}
\newcommand{\CC} {{\cal P}_{cl}}
\newcommand{\CQ} {{\cal P}_{Q}}
\newcommand{\CA} {{\cal P}_{\alpha}}
\newcommand{\CG} {{\cal P}_{\gamma}}

\newcommand{\AH} {{\hat A}}
\newcommand{\XH} {{\hat X}}
\newcommand{\CXH} {{\hat {\cal P}}_{X}}
\newcommand{\JCH} {{\hat J}_{cl}}
\newcommand{\QH}  {{\hat Q}}
\newcommand{\JAH} {{\hat J}_{\alpha}}
\newcommand{\JBH} {{\hat J}_{\beta}}
\newcommand{\JGH} {{\hat J}_{\gamma}}
\newcommand{\MA} {m_{\alpha}}
\newcommand{\MG} {m_{\gamma}}

\newcommand{\JEH} {{\hat J}_{\eta}}

\newcommand{\JH} {{\hat J}}

\begin{document}

\reversemarginpar
\tighten

\title{Quantum spectrum for a Kerr-Newman black hole}

\author{Gilad Gour\thanks{E-mail:~gilgour@phys.ualberta.ca}
and A.J.M. Medved\thanks{E-mail:~amedved@phys.ualberta.ca}}

\address{
Department of Physics and Theoretical Physics Institute\\
University of Alberta\\
Edmonton, Canada T6G-2J1\\}

\maketitle

\begin{abstract}
In this paper, we consider the quantum area spectrum
for a rotating and charged (Kerr-Newman)
black hole.  Generalizing a  recent 
study on  Kerr black holes 
 (which was inspired by the static-black hole formalism
of  Barvinsky, Das and Kunstatter),
 we show that the quantized area operator
can be expressed in terms of three quantum numbers
(roughly related to the mass, charge and spin sectors).
More precisely, we find that $A=8\pi\hbar[n+{1\over 2}+{p_1\over 2}+p_2]$,
where $n$, $p_1$ and $p_2$ are strictly non-negative integers.
In this way, we are able to confirm
a uniformly spaced spectrum  even for a fully
general Kerr-Newman black hole. Along the way, we  derive
certain selection rules and use these to demonstrate
that, in spite of appearances, the charge and spin spectra 
are not completely independent.
\end{abstract}

\section{Introduction}

\par
There has been significant interest, over the last 
 decade,  in the
subject known as black hole spectroscopy.
(For an overview, see \cite{BEKX}.)
The general concept is that a  black hole observable,
in particular the surface area of the horizon, should
be quantized in terms of a discretely changing quantum
number (or numbers).
That this should be the case was first advocated
by Bekenstein \cite{BEK2}, not long after the
realization of black holes as thermodynamic systems
 \cite{BEK,HAW}.
 The main point of Bekenstein's
argument is that, for a slowly evolving black hole, the horizon  area 
($A$) behaves as an adiabatic invariant 
\cite{BEK3}. The significance of this property follows from
Ehrenfest's principle, which indicates that
a classical adiabatic invariant corresponds to a quantum
observable with a discrete spectrum.

\par 
Bekenstein  went on to  suggest
 that the quantum area spectrum  should
be evenly spaced with increments of size $\epsilon \hbar$ \cite{BEK2,MUK},
%following form for the  area spectrum of a static
%and uncharged black hole 
% \cite{BEK2,MUK}:
%\be
%A=\epsilon\hbar  n, \quad\quad\quad n=0,1,2,...,
%\label{0.2}
%\ee
where $\epsilon$ is a numerical factor of the order unity and
$\hbar$ is the Planck constant.  (Here and throughout,
we restrict the discussion to a four-dimensional spacetime
and set all other fundamental constants to unity.)
%The most important feature in this
%expression is 
% the uniform spacing  between the
%levels. 
This discreteness can be viewed as a consequence of  the
uncertainty principle, which tells us that  a quantum point particle
can, at best, only be  localized to  a single  Compton length.
%Significantly,  this non-locality  implies a minimal increase in the
%horizon area, which turns out to be
% $(\Delta A)_{min}=\epsilon \hbar$ \cite{BEKX}.

\par
Much of the subsequent interest in black hole spectroscopy
has centered around the idea that such 
 heuristic arguments 
  can somehow be substantiated by more rigorous means.
There has, indeed, been significant progress in studies
 along this
line (see \cite{X2} for references); for instance, 
the algebraic approach to black hole quantization, as
developed by Bekenstein and one of the current 
authors~\cite{X1,X2,X3}.
In this approach, starting
with very elementary assumptions and then exploiting symmetries, one  
can obtain the algebra of the relevant operators of the black
hole. 
 Bekenstein has used this methodology to provide  
 a more rigorous proof~\cite{X1} for the equally spaced
area spectrum. In the subsequent analysis~\cite{X2,X3}, 
the neutral, non-rotating 
black hole observables have been constructed by subjecting a pair of
creation operators, or ``building blocks'', to a 
simple algebra. This construction reproduces the evenly spaced area 
spectrum, provides a rigorous proof that the $n$-th area eigenvalue 
is exponentially degenerate and predicts that a logarithmic correction
term, $-3/2\log A$, should be added to the Bekenstein-Hawking 
entropy \cite{BEK,HAW}.  
 
\par
Another such program, which is particularly
relevant to the current paper, was 
initiated   by Barvinsky and Kunstatter \cite{BG}.
To  summarize,
these authors expressed the
black hole dynamics in terms of a  reduced phase space
(the reduction comes by way of a minisuperspace type of approximation),
which they were then able to quantize.
For a static and  uncharged black hole (which was the
focus of this seminal paper),
the reduced phase space consists of only the black hole mass 
observable and its canonical conjugate \cite{KUN,KUC}.\footnote{Although
the simplicity of this picture may appear counter-intuitive,
it actually follows quite naturally from either
 Birkhoff's theorem \cite{WAL} or
the  ``no-hair'' principles of black holes \cite{NH}.}
With one justifiable assumption  - namely, the
conjugate to the mass  is identified
with a periodicity which coincides with
that of Euclidean time \cite{GH} - the authors
reproduced a uniformly spaced area spectrum such that
$\epsilon=8\pi$ \cite{BG}. 
(They also found  a zero-point contribution of $4\pi \hbar$,
which in no way undermines the original Bekenstein proposal.)

\par
The methodology of \cite{BG} was later extended
for the highly  non-trivial inclusions of non-vanishing charge
(by Barvinsky, Das and Kunstatter \cite{BDG,BDGX}) and
spin
(by the current authors \cite{GJ}).
 In the former, charged case, the reduced phase
space  consists of the two relevant observables 
(the mass, $M$, and the charge, $Q$) and their respective conjugates
\cite{LMG}. Imposing the same periodicity condition as discussed above,
Barvinsky {\it et al} found  an  area spectrum of the
following form \cite{BDG}:
\be
A-A_{ext}(Q)= 8\pi \hbar \left(n+  {1\over 2}\right), 
\quad\quad\quad n=0,1,2,...,
\label{0.3}
\ee
where  $A_{ext}=A_{ext}(Q)$ is the extremal value of the horizon 
area.\footnote{A
charged and/or rotating black hole typically has a pair of
distinct horizons,  with the point of coincidence defining
the  extremal  horizon. Note that, in this paper,
 an unqualified $A$ will always signify the
area of the outermost horizon.}
(It should be kept in mind that $A_{ext}$ represents
 a lower bound on the horizon area of
a classical black hole. Note, however,
that the zero-point term in Eq.(\ref{0.3}) prohibits
the quantum black hole from actually  approaching this extremal
value.)  After quantizing the charge sector of the theory,
Barvinsky {\it et al} finally obtained \cite{BDG}
\be
A=8\pi \hbar \left(n+ {p\over 2} +{1\over 2}\right), 
\quad\quad\quad n,p=0,1,2,...,
\label{0.4}
\ee
where  $p$ is related to the
black hole charge via $Q^2=\hbar p$.

\par
The latter, rotating case was complicated by 
the lack of a concrete example for the  reduced phase
of a spinning black hole. Nevertheless, we argued 
for the existence of such a reduced space by appealing
to the ``no-hair'' principles of black holes \cite{NH}.
With this assumption and the usual periodicity
constraint, we were able to deduce an area spectrum
of the following form \cite{GJ}:
\be
A-A_{ext}(\JC)= 8\pi \hbar \left(n+  {1\over 2}\right),
\quad\quad\quad n=0,1,2,...,
\label{0.5}
\ee
where $\JC$ is a rotation operator that is
related to but distinct from the angular momentum
of the black hole.\footnote{To emphasize
this distinction, we always use, in both the prior
\cite{GJ} and current paper, 
the subscript $cl$
for {\it classical}. This is, however, somewhat
misleading, as $\JC$ is ultimately
elevated to the status of a quantum operator at some
point in the analysis.} 
After quantizing the spin sector, we found that the
spectrum (\ref{0.5}) could be re-expressed as follows \cite{GJ}:
\be
A=8\pi \hbar \left(n+ m +{1\over 2}\right), 
\quad\quad\quad n,m=0,1,2,...,
\label{0.6}
\ee
with the rotation operator having been    quantized according to
$\JC=\hbar m$.  It should be noted that, in the limit of
very large spin or $m>>1$, $m\sim j$ where
 $j$ is the angular-momentum eigenvalue of
the black hole.
\par
In our prior paper,
we purposefully neglected any consideration of
charge so as to stress the technical issues
that are indigenous to the  case of a rotating black hole.
In the current paper, we rectify this omission
and extend the methodology to a black hole
with both charge or spin; that is, we determine
the area spectrum of a fully general Kerr-Newman
black hole. Note that, to avoid superfluous
repetition, we will go over some steps rather
quickly.  For a better understanding of the
subtleties  of this procedure,
the reader is referred to the earlier
paper \cite{GJ}. Also, let us take this moment to
emphasize that {\it only} two (well-motivated) assumptions go
into our quantization procedure: 
 a rotating black hole can be described, in analogy
to studies on static black holes \cite{KUN,KUC,LMG},
by a relatively simple form of reduced phase space 
{\it and} the conjugate to the mass is periodic 
in accordance with Euclidean considerations \cite{KUC,GH,BG}.

\par
It should be noted that, in a recent study of interest,
 Makela {\it et al} have similarly considered
the area spectrum of the Kerr-Newman black hole \cite{MRLP}.
The approach of these authors (also see~\cite{LM})  is based on formulating
a Schrodinger-like equation for the black hole
observables and  quantizing this equation by way of
 WKB  techniques.
Their form of the area spectrum differs somewhat with 
what we eventually derive here; 
however, a direct comparison
is highly non-trivial due to a fundamental
distinction in what quantity is precisely being quantized.
More specifically, Makela {\it et al} quantized 
  $A + A_{-}$ (where
$A_{-}$ is the area of the inner black hole horizon)
as opposed to $A-A_{ext}$ ({\it cf}, Eqs.(\ref{0.3},\ref{0.5})).

\par
The remainder of the paper is organized as follows. In 
 Section 2, we consider 
 a Kerr-Newman black hole at the classical level,  with
particular emphasis 
on the (conjectured) reduced phase space. 
The quantization 
of this  phase space, as detailed in Section 3,  leads 
to an area spectrum
that is formally analogous to   Eqs.(\ref{0.3},\ref{0.5}).
In Section 4, we focus on the charge and spin sector,
and develop a means by which this can be suitably quantized.
 After implementing certain {\it selection rules}, we are able
to show that the
area spectrum is, indeed, uniformly spaced,
even for a fully general Kerr-Newman solution.
The final section contains a brief summary
and some further discussion on the selection rules.

\section{Classical Analysis}

We begin by considering the   
Kerr-Newman black hole, which may be regarded as
the most general   solution of the vacuum Einstein equations.
Because of the ``no-hair'' principles \cite{NH},
it is expected that an external observer
can describe
 this system
 in terms of a few macroscopic parameters:
 the black hole mass, $M$,  charge, $Q$, and an  angular
momentum, $\vec{J}_{cl}$.  
Furthermore,
the  first law of black hole mechanics \cite{BEK,HAW}
allows us to relate these quantities in the following manner:
\be
dM={\kappa\over 8\pi}dA+ \Phi dQ +\Omega d\JC.
\label{1}
\ee
Here, 
$A$ is the (outermost) horizon area, $\kappa$ is
the surface gravity at this horizon, $\Phi$ is the electrostatic
potential,
$\Omega$ is the angular velocity,  
and $\JC=|\vec{J}_{cl}|$ is the magnitude
of the  angular-momentum vector.

\par
The  thermodynamic properties of a Kerr-Newman
black hole are explicitly known \cite{WAL} and  expressible as follows:
\be
A=8\pi M\left[M -{Q^2\over 2M}+\sqrt{M^2-Q^2-{\JC^2\over M^2}}\right]
\label{2}
\ee
or equivalently
\be
M^2={A\over 16\pi}+4\pi{\JC^2\over A}+{Q^2\over 2}
+{1\over \pi}{Q^4\over A},
\label{3}
\ee
and
\be
\kappa= {1\over 4M}-16\pi^2{\JC^2\over MA^2}-4{Q^4\over
MA^2},
\label{4}
\ee
\be
\Phi= {Q\over 2M}\left[1+{4 Q^2\over \pi A}\right],
\label{4.5}
\ee
\be
\Omega= 4\pi{\JC\over MA}.
\label{5}
\ee

\par
Generalizing the philosophy of \cite{GJ},
we will assume that the Kerr-Newman black hole can
be  dynamically expressed, at the classical level, in terms of
 a reduced phase
space consisting of the physical observables
and their  canonical conjugates.
(For further justification, also see \cite{MRLP}.)
More specifically, we propose that the phase
space can be described by 
the following  set of observables:
\be
A , Q ,  \JA , J_{\beta} , \JG ,
\label{6}
\ee
along with their respective conjugates
\be
\PAA , \PQ, \alpha , \beta , \gamma .
\label{7}
\ee
In this formulation, $\JA$, $J_{\beta}$, and $\JG$ represent   
the {\it Euler} components of the angular momentum \cite{EUL} and their
conjugates, $\alpha$, $\beta$ and $\gamma$,
are  the corresponding
Euler angles. For future reference, note that \cite{EUL}
\be
\JX= -\cos\alpha\cot\beta\JA-\sin\alpha\JB
     +{\cos\alpha\over\sin\beta}\JG,
\label{11}
\ee
\be
\JY= -\sin\alpha\cot\beta\JA+\cos\alpha\JB
     +{\sin\alpha\over\sin\beta}\JG,
\label{12}
\ee
\be
\JZ=\JG,
\label{13}
\ee
where $\JX$, {\it etc.} are the Cartesian components
of the angular momentum.

\par
Intuitive considerations suggest that
the horizon area is invariant under rotations of its spin 
and under gauge transformations.
It is thus follows that Eqs.(\ref{6},\ref{7}) form a
 set of {\it generalized commuting} coordinates and their conjugates.
(In the algebraic approach~\cite{X1}, one similarly starts 
with the area, charge 
and angular momentum as the initial commuting observables.)  
On the other hand, we ultimately want to work 
with the mass, $M$, rather than the area, $A$
(this allows us to exploit the periodicity
of the mass conjugate, as noted
in Section 1), and it can be shown that \cite{GJ}
\be
\lbrace M,J_{\beta} \rbrace \neq 0.
\label{new1}
\ee
(Here, $\lbrace\quad,\quad \rbrace$  denotes a commutator
or Poisson bracket in the Dirac sense \cite{DIR}.)
We can, however, circumvent this awkward situation
by replacing $J_{\beta}$ with $\JC$.

\par
One can see the relevance of the proposed  ``switch''  by
inspecting the explicit form of  $\JC$:
\bea
\JC^2&=&\JX^2+\JY^2+\JZ^2
\nonumber \\
&=& {1\over \sin^2\beta}\left[\JA^2+\JG^2-2\cos\beta
\JA\JB\right]+\JB^2,
\label{14}
\eea
where we have applied Eqs.(\ref{11}-\ref{13}) and
treated  $\JA$, $J_{\beta}$, $\JG$  as classical
or  non-operating quantities. It is, in fact, the  presence 
of $\beta$ (but neither $\alpha$ nor $\gamma$) in 
the above expression that makes $J_{\beta}$
a rather poor choice in constructing the phase space.

\par 
With prompting from the above discussion, we now consider
a new set of observables:
\be
M=M(A,Q,\JC), Q ,  \JC , \JA , \JG 
\label{15}
\ee
and denote their respective conjugates as follows:
\be
\IM, \IQ ,  \IC, \IA, \IG.
\label{16}
\ee
The transformation from Eqs.(\ref{6},\ref{7})
into Eqs.(\ref{15},\ref{16}) is canonical if
\be
\lbrace M,\IM \rbrace = \lbrace Q, \IQ \rbrace =
 \lbrace \JC,\IC \rbrace
=\lbrace \JA,\IA \rbrace = \lbrace \JG,\IG \rbrace
=1, 
\label{17}
\ee
\be
\lbrace all\quad other\quad combinations \rbrace=0,
\label{18}
\ee
where the derivatives are taken with respect to
the original set of generalized coordinates  (\ref{6},\ref{7}).

\par
Applying some straightforward but lengthy calculations,
one can readily verify that the proposed transformation
is canonical provided that
\be
\IM={8\pi\over \kappa} \PAA,
\label{20}
\ee
\be
\IQ=\PQ-{8\pi\over \kappa}\Phi \PAA,
\label{20.5}
\ee
\be
\IC=\PC-{8\pi\over\kappa}\Omega\PAA,
\label{21}
\ee
\be
\IA=\PA,
\label{22}
\ee
\be
\IG=\PG.
\label{23}
\ee
Note that $\PC$ is defined by  first making a canonical transformation
from Eq.(\ref{6},\ref{7}) to the same set but with $\JC$ ($\PC$)
replacing $J_{\beta}$ ($\beta$).

\section{Quantum Analysis}

\par
To proceed with a suitable process of quantization,
in the manner originally advocated by Barvinsky and Kunstatter \cite{BG},
the following condition of periodicity is required:
\be
\IM\sim\IM+{2\pi\over\kappa}.
\label{24}
\ee
Although technically  an assumption, this condition
has a well-justified pedigree that follows
from the known periodicity of Euclidean time \cite{GH} 
and the identification of $\IM$ with
a measure of Schwarzschild-like time \cite{KUC}.
(Consult \cite{BG,BDG,BDGX,GJ} for further discussion.)

\par
Following the prescribed program, we now proceed by introducing
a  pair of variables that directly incorporate the 
periodic nature of $\IM$:
\be
X=\sqrt{\hbar B(M,Q,\JC,\JA,\JG)\over\pi}\cos(\kappa\IM),
\label{25}
\ee
\be
\CX=\sqrt{\hbar B(M,Q,\JC,\JA,\JG)\over\pi}\sin(\kappa\IM).
\label{26}
\ee
The yet-to-be-specified function $B$ will be partially fixed via the
  constraint that 
Eqs.(\ref{15},\ref{16}) transform {\it canonically} into
the set of observables
\be
X, Q,\JC,\JA,\JG
\label{27}
\ee
and respective conjugates 
\be
\CX,\CQ, \CC,\CA,\CG.
\label{28}
\ee

\par
Let us therefore consider the
following necessary and sufficient condition for
a canonical transformation:
\bea
\CX\delta X + \CQ\delta Q + \CC\delta\JC + \CA\delta\JA +\CG\delta\JG
=\nonumber \\
\IM\delta M + \IQ\delta Q + \IC\delta\JC +\IA\delta\JA +\IG\delta\JG.
\label{29}
\eea
Up to a total variation, one finds that
\be
\CX\delta X={\hbar\kappa\IM\over 2\pi}\left[
{\dd B\over \dd M}\delta M +
{\dd B\over \dd Q}\delta Q + 
{\dd B\over \dd \JC}\delta \JC + 
{\dd B\over \dd \JA}\delta \JA + 
{\dd B\over \dd \JG}\delta \JG \right].
\label{30}
\ee 
Substituting Eq.(\ref{30}) into Eq.(\ref{29}), we
promptly obtain the following pivotal result:
\be
{\dd B\over\dd M}= {2\pi\over\hbar\kappa}.
\label{31}
\ee
For future reference, we also have
\be
{\dd B\over\dd Q}= {2\pi\over\hbar\kappa\IM}\left(\IQ-\CQ\right),
\label{31.5}
\ee
\be
{\dd B\over\dd \JC}= {2\pi\over\hbar\kappa\IM}\left(\IC-\CC\right)
\label{32}
\ee
and analogous expressions involving  $\CA$ and $\CG$.

\par
It is  instructive to compare Eq.(\ref{31}) with
the first law, Eq.(\ref{1}),
indicating that $\dd A/\dd M=4\hbar \dd B/\dd M$. This
suggestive result directly implies the following:
\be
B(M,Q,\JC,\JA,\JG)={1\over 4\hbar}A(M,Q,\JC)+F(Q,\JC,\JA,\JG),
\label{35}
\ee
where $F$ is an essentially arbitrary function
of the charge and angular momentum. That is to say, for any
well-behaved choice of $F$, one will always be
able to find expressions for $\CQ$, $\CC$, $\CA$ and 
$\CG$ that maintain a canonical transformation.

\par
In spite of this apparent freedom in  $F$,
we are  able to fix this function by way of the following 
argument.
Let us first point out that the (outer) horizon area
of a Kerr-Newman black hole is bounded from below 
- at least classically - 
by its extremal value \cite{WAL};
that is:\footnote{This extremal area can be obtained
by constraining the mass observable so
that the square-root argument in Eq.(\ref{2}) is
exactly vanishing.}
\be
A \geq A_{ext}= 4\pi\sqrt{Q^4+ 4\JC^2}.
\label{36}
\ee 
As elaborated on in the  related works \cite{BG,BDG,BDGX,GJ}, 
it is most  convenient
if $F$ is chosen so that Eq.(\ref{36}) translates
into the bound $B\geq 0$.  On this basis,
we can unambiguously  set $F=-A_{ext}/4\hbar$ and obtain
\be
 B= {1\over 4\hbar}\left[
A(M,Q,\JC)- 4\pi\sqrt{Q^4 +4 \JC^2}\right] .
\label{37}
\ee

\par
Let us now reconsider Eqs.(\ref{25},\ref{26}), which
can be squared and summed to yield $\hbar B=\pi(X^2+\CX^2)$.
Incorporating this finding into Eq.(\ref{37}),
we have
\be
X^2+\CX^2 = {1\over 4\pi}\left[A(M,Q,\JC)-
4\pi\sqrt{Q^4+4 \JC^2}\right]
\geq 0.
\label{38}
\ee
It is especially relevant that   the mass and its conjugate,
$M$ and $\IM$, have been mapped into a {\it complete} two-dimensional plane,
$X$ and $\CX$. Any other choice of $F$ would have left
a ``hole'' in this plane and unnecessarily complicated 
the impending process of quantization. 

\par
Next, let us elevate the classically defined quantities 
to   quantum operators (denoted by ``hats''). 
Eq.(\ref{38}) then takes
the following form:
\be
{\hbar\over 2\pi}{\hat B}\equiv
{1\over 8\pi}\left[\AH-4\pi\sqrt{\QH^4 + 4 \JCH^2}\right]
={\XH^2\over 2}+{\CXH^2\over 2}.
\label{39}
\ee
Given that the domain of $\XH$ and $\CXH$ is a complete two-dimensional
plane, the spectrum of these operators is  trivially
that of a harmonic oscillator. Hence, we can write
\be
 B_n =
 2\pi\left[n+{1\over 2}\right],
\quad\quad\quad n=0,1,2,....
\label{40}
\ee
where $B_n$ are the eigenstates of the operator
${\hat B}$. Keep in mind that $B_n$ is essentially
a measure of the deviation of the horizon area
from extremality.

\par
It is interesting to note that,
by virtue of the zero-point term in Eq.(\ref{40}),
 quantum fluctuations
will  inhibit the  black hole
from ever reaching a precise state of extremality.
A similar observation has been made  for both charged (but non-rotating)
 \cite{BDG} and rotating (but uncharged) \cite{GJ} black holes.  

\section{Charge and Spin Sector}

\par
Our task is not yet complete, as we still require
that the spectra for $\AH$ and 
$\AH_{ext}=4\pi\sqrt{\QH^4+\JCH^2}$
be explicitly separated.
(Note that, for this purpose, a complete separation
of the spectra for $\JCH$ and $\QH$ is not necessarily required.)
It turns out that this objective can readily be accomplished 
by way of some straightforward arguments.

At a first glance, the quantization of $\AH_{ext}$ appears to be 
 a trivial process; inasmuch as  $\hat{Q}$ is the generator 
of $U(1)$ gauge transformations, its spectrum of eigenvalues ($Q$) must
certainly  be of the form
\be
Q=em_{1}, \quad\quad\quad m_{1}=0,\pm1,\pm2,..., 
\label{01}
\ee
where $e$ is the fundamental unit of electrostatic charge. Furthermore, 
 the quantum limit of the angular-momentum operator 
({\it cf}, Eq.(\ref{14})),
\be
\JCH^2 
= {1\over \sin^2\beta}\left[\JAH^2+\JGH^2-2\cos\beta
\JAH\JBH\right]+\JBH^2,
\ee
has been shown~\cite{GJ}  to have the following discrete set of eigenvalues
 ($J_{cl}$):
\be
J_{cl}=\hbar m_{2}, \quad\quad\quad  m_{2}=0,1,2,....
\label{02}
\ee
Therefore, by substituting Eqs.(\ref{01},\ref{02}) (and also Eq.(\ref{40}))
 into 
Eq.(\ref{39}), we find that the area spectrum can be expressed
as follows:
\be
A_{n,m_{1},m_{2}}=8\pi\hbar\left[n+{1\over 2}\right]+
{1\over 2}\sqrt{e^{4}m_{1}^{4}+4\hbar ^{2}m_{2}^{2}}.
\label{03}
\ee
Given this form of the  spectrum,
it is not at all obvious  that the  levels
could, in general, be  evenly spaced.
 Nonetheless,
by utilizing the periodicity of $\Pi_M$ to impose
{\it selection rules} on   $m_{1}$ and $m_{2}$, 
we will  demonstrate below that  a uniformly
spaced spectrum is, indeed, consistently realized.  

\par
With the above in mind,
let us first consider the quasi-charge sector.\footnote{We use
this ``quasi'' terminology because the charge and spin sectors
do not, in general, separate completely.}
It is useful
to  recall Eq.(\ref{31.5}):
\be
\CQ=\IQ+\Phi\IM
+{\kappa\over 8\pi}\IM{\dd A_{ext}\over\dd Q},
\label{501}
\ee
where the first law of black hole mechanics (\ref{1})
and the precise forms of $B$ (\ref{35})
and $F=-A_{ext}/4\hbar$ have also  been  applied.
For sake of clarity, let us further re-express this
relation as follows:
\be
\CQ=\chi_1+ {\theta\over 8\pi}{\dd A_{ext}\over\dd Q},
\label{502}
\ee
where $\chi_1\equiv\IQ+\Phi\IM$ and $\theta\equiv\kappa\IM$.
It should be kept in mind that $\theta$ is an angle
({\it i.e.}, has a periodicity of exactly $2\pi$); {\it cf},
Eq.(\ref{24}). Note that, for a non-rotating black hole,
it has been shown \cite{BDG} that
the variable $\chi_1$ is, up to a dimensional factor, 
also an angular quantity. However, this need not be the case
when the spin is ``turned on'', and so
we will not apply (nor require) this result.

\par
Now consider that, in the coordinate representation
with $\QH=-i\hbar\dd/\dd\CQ$, the wavefunctions for
the charge eigenstates  take the form
\be
\Psi_{Q}(\PQ)\sim \exp\left[{iQ\CQ\over \hbar}\right]
\label{503}
\ee
and  we can, therefore,  
make the following identification:
\be
{Q\CQ\over \hbar}\sim {Q\CQ\over \hbar}+ 2\pi n_1,
\label{504}
\ee
where $n_1$ is an arbitrary integer.

\par
Let us next consider the implication of Eqs.(\ref{502}) and
(\ref{504}) when taken together. Holding $\chi_1$ 
constant,\footnote{Here, we are treating $\chi_1$ and
$\theta$ as independent variables. One might be concerned
that both depend on the conjugate $\IM$; however,
$\chi_1$ also depends on a variable, $\IQ$,
that is clearly independent of $\IM$.  Hence,
we argue that $\chi_1$ can be held constant
without loss of generality.}
we are able to deduce that
\be
\theta{Q\over 8\pi\hbar}{\dd A_{ext}\over\dd Q}
\sim \theta {Q\over 8\pi\hbar}{\dd A_{ext}\over\dd Q}
+2\pi n_2,
\label{505}
\ee
where $n_2$ is another arbitrary integer.
That is, the quantity on the left-hand side
must necessarily be an angle. However, 
$\theta$ is, by hypothesis, itself an angle, and so
we can write
\be
{Q\over 8\pi\hbar}{\dd A_{ext}\over\dd Q}=p_1,
\label{506}
\ee
where $p_1$ is yet another integer
(which is manifestly non-negative, as can be
seen by inspecting the left-hand side).

\par
We can make the above expression more explicit
by utilizing Eq.(\ref{36}) to obtain
\be
{1\over \hbar}{Q^4\over\sqrt{Q^4+4\JC^2}}=p_1, \quad\quad\quad p_1=0,1,2,....
\label{507}
\ee
This quantization condition will be referred to as {\it selection
rule} $no. 1$.

\par
Let us next examine the quasi-spin sector,
beginning with the appropriately
revised form of Eq.(\ref{32}):
\be
\CC= \chi_2
+{\theta\over 8\pi}{\dd A_{ext}\over\dd \JC},
\label{508}
\ee
where $\chi_2\equiv\IC+\Omega\IM$.
Because of the obvious symmetry between this
relation and its quasi-charge sector analogue
(\ref{502}), it is clear that the current quantization
procedure will closely follow the prior analysis.
Repeating the steps, as outlined above, for
the current case, we find that 
\be
{\JC\over 8\pi\hbar}{\dd A_{ext}\over\dd \JC}=p_2,
\label{509}
\ee
where  $p_2$ is strictly a non-negative
integer.

\par
Substituting Eq.(\ref{36}) into the above relation,
we obtain  {\it selection rule} $no. 2$:
\be
{1\over \hbar}{2\JC^2\over\sqrt{Q^4+4\JC^2}}=p_2, 
\quad\quad\quad p_2=0,1,2,....
\label{510}
\ee

\par
Next, we consider an appropriate linear combination
of the two selection rules (\ref{507},\ref{510}).  
Multiplying $no. 2$  by two and adding this to
$no. 1$, we have (after some trivial
manipulations) 
\be
\sqrt{Q^4+4\JC^2}=\hbar\left[p_1 +2p_2\right],
\quad\quad\quad p_1,p_2=0,1,2,....
\label{514}
\ee

\par
Let us now recall and appropriately rearrange Eq.(\ref{37}):
\be
A= 4\hbar B + 4\pi\sqrt{Q^4 +4 \JC^2}.
\label{555}
\ee
Quantizing this relation and then incorporating Eqs.(\ref{514},\ref{40}), 
we can now write the
area spectrum 
in the following elegant form:
\be
A_{n,p_1,p_2}=8\pi\hbar\left[n+{1\over 2}+ {p_1\over 2}+p_2\right]
 \quad\quad\quad n,p_1,p_2=0,1,2,....
\label{516}
\ee
Hence, we have realized an evenly spaced spectrum
for the area of a fully general Kerr-Newman black hole.

\par
It should be kept in mind that, strictly speaking, the quantum numbers $p_1$ 
and $p_2$ are {\it not} independent parameters. Rather, these
integers  are related by way of the selection rules; so that, 
once $p_1$ has been fixed, $p_2$ will be restricted to
a limited  set of allowable values and
{\it vice versa}. We will elaborate on this point in
the section to follow. 

\section{Concluding Discussion: Selection Rules}

In summary, we have considered the quantum area
spectrum of a Kerr-Newman (rotating and charged) black
hole. Extending the methodology of Barvinsky {\it et al} 
for a static system \cite{BG,BDG}, as well as
a recent treatment on
Kerr  black holes \cite{GJ}, we
have obtained  an explicit form for
the area spectrum in terms of   three integer-valued
quantum numbers. Moreover, the spectrum was shown
to be uniformly
spaced, in direct compliance with the heuristic arguments
of Bekenstein~\cite{BEK2,MUK,BEKX} and the algebraic 
approach~\cite{X1,X2,X3}. 

\par
Let us re-emphasize that some conjectural, although well-motivated
inputs were used in attaining this result.  First of all,
it was assumed that a reduced phase space description
exists for a  fully general Kerr-Newman black hole.
Secondly, we  utilized a periodicity
condition on the conjugate to the mass that was first
proposed by Barvinsky and Kunstatter~\cite{BG}.
Nonetheless, we feel that the elegance of the
resulting spectrum only strengthens 
our convictions with regard to the use of
such inputs. We expect, however, to more rigorously address
these issues at a future time.

\par
Finally, let us more closely examine the implications
of the selection rules on the charge and spin spectra.
For illustrative purposes, we will first focus on the case
of vanishing spin.  Then 
Eqs.(\ref{01},\ref{507}) require
\be
 \eta\equiv {e^2 \over \hbar}= {p_1\over m_1^2}; 
\label{602}
\ee
that is, this ratio of fundamental constants, $\eta$,
is constrained to be a rational, constant number. 
It follows that  $\eta$  can always 
 be expressed in the following manner:
\be
\eta ={a\over b},
\label{04}
\ee
where $a$ and $b$ are mutually prime 
({\it i.e.},  non-divisible) integers.  Therefore, 
Eq.~(\ref{602}) rules out many eigenvalues of $\hat{Q}$
and only the following are allowed:
\be
Q=e m_{1},\quad\quad\quad{\rm where}\quad\quad\quad m_1^2 \; {\rm mod} \; b =0.
\label{05}
\ee
That is to say, $m_1^2$ must be divisible by $b$.

\par
For the case of vanishing charge but non-vanishing spin, 
the results are substantially different, as
Eqs.(\ref{02},\ref{510}) do not provide any new information. Rather,
we only have the trivial outcome of $m_2=p_2$
for a neutral black hole.

\par
Next, moving on to the fully general case,
we find (after some manipulations) that the  selection 
rules~(\ref{507},\ref{510}) now imply a pair of constraints:
\be
\eta^2 m_{1}^{4}=p_{1}(p_{1}+2p_{2})\quad\quad\quad{\rm and}\quad\quad\quad
2 m_{2}^{2}=p_{2}(p_{1}+2p_{2}).
\label{603}
\ee
Since $\eta$ is the same rational number given by Eq.(\ref{04}),
the charge sector is still quantized according to Eq.~(\ref{05}).
Furthermore, Eq.~(\ref{603}) shows that, for any definite value
of $m_{1}$,  $m_{2}$  can only take on a limited set of allowable values
and
{\it vice versa}. To put it another way: as a consequence of
the selection rules, the  charge and  
angular-momentum spectra are not completely independent.
 
\par
It has been suggested by Barvinsky {\it et al} \cite{BDG}
that the above type of analysis can also be used
to fix the fundamental constants of nature, $\hbar$ and
$e$, by a Coleman-like ``big-fix'' mechanism \cite{COL}.
Certainly, the ratio $e^2/\hbar$  must necessarily
be a rational number ({\it cf}, Eq.(\ref{602}))
independently of any other considerations.
However, it remains unclear if this outcome
is an artifact of an intrinsically semi-classical
framework or a manifestation of some deep,  fundamental
principle of quantum gravity. We can only
hope that future investigations can shed some
light on this intriguing question.

\section{Acknowledgments}

The authors graciously thank V.P. Frolov for
helpful conversations and J.D. Bekenstein and G. Kunstatter
for  enlightening correspondences. GG is also grateful for
the Killam Trust for its financial support.

\end{document}